\documentclass[a4paper,11pt]{article}
\pdfoutput=1 

\usepackage{jheppub} 

\usepackage[T1]{fontenc} 
\usepackage{comment}
\usepackage{tikz}
\usepackage{amsmath,amssymb}

\title{Towards higher-spin holography in flat space}

\author{Dmitry Ponomarev}

\affiliation{Institute for Theoretical and Mathematical Physics,\\
Lomonosov Moscow State University, Lomonosovsky avenue, Moscow, 119991, Russia}
\affiliation{I.E. Tamm Theory Department, Lebedev Physical Institute,\\
 Leninsky avenue, Moscow, 119991, Russia}

\emailAdd{ponomarev@lpi.ru}

\abstract{
We study the chiral flat space higher-spin algebra, which is  the global symmetry algebra of the chiral higher-spin theory  in the 4d Minkowski space. We find that it can be constructed as the universal enveloping algebra of a certain chiral deformation of the Poincar\'{e} algebra quotiented by a set of quadratic identities. These identities allow us to identify a representation of the latter algebra, which by analogy with the AdS space higher-spin holography, we interpret as the flat space singleton representation. We provide two explicit realisations of this singleton representation -- in terms of $sl(2,\mathbb{C})$ spinors and in terms of  oscillator-like variables -- as well as briefly discuss its properties. 
}

\begin{document} 
\maketitle
\flushbottom

\section{Introduction}

In recent years significant interest was drawn to flat space holography. For reviews on the subject we refer the reader to   \cite{Strominger:2017zoo,Raclariu:2021zjz,Pasterski:2021rjz,Prema:2021sjp,Pasterski:2021raf,McLoughlin:2022ljp}. There are different approaches to  flat space holography, but all of them acknowledge the central role of the (extended) BMS symmetry \cite{Bondi:1962px,Sachs:1962wk} -- or, equivalently, the conformal Carroll symmetry \cite{Duval:2014uva} -- as the symmetry underlying the dual boundary dynamics \cite{Barnich:2009se,Strominger:2013lka,Strominger:2013jfa,Barnich:2013sxa,He:2014cra,Campiglia:2015qka,Campoleoni:2017mbt,Donnay:2020guq,Campoleoni:2020ejn}. Similarly to the AdS/CFT correspondence, in flat holography one reinterprets  flat-space scattering amplitudes as  correlators in the dual theory, which, upon recasting them in a boost eigenstate basis  \cite{Pasterski:2016qvg,Pasterski:2017kqt,Banerjee:2018gce,Pasterski:2020pdk}, do manifest properties of some exotic conformal field theories \cite{Pasterski:2017ylz,Stieberger:2018edy,Fan:2019emx,Nandan:2019jas,Pate:2019lpp,Banerjee:2020kaa,Jiang:2021csc,Adamo:2021zpw,Donnay:2022aba,Bagchi:2022emh} -- living either on the celestial sphere or on the Minkowski space boundary. Despite significant progress in this direction, no explicit flat space holographically dual pairs have been constructed yet\footnote{See, however, recent works \cite{Costello:2022jpg,Stieberger:2022zyk,Rosso:2022tsv}.}. 

One simple reason why constructing holographically dual pairs with  flat space bulk is problematic is the qualitative difference in the representation theory of the AdS and the Minkowski space isometries. To be more precise, for the case of a $d$-dimensional bulk, the boundary theory is $(d-1)$-dimensional and, accordingly, the boundary on-shell fields have the Gelfand-Kirillov (GK) dimension equal to $d-2$\footnote{If a representation space is carried by functions of $k$ independent variables, its GK dimension is equal to $k$. This is the standard way to characterise the size of an infinite-dimensional representation. The GK dimension of an on-shell theory in $d-1$ dimensions is $d-2$, because the on-shell condition removes dependence of a field on one variable, which  otherwise is given by an arbitrary function of all $d-1$ space-time coordinates.}. What makes  flat space holography problematic is that  the Poincar\'{e} algebra $iso(d-1,1)$ -- which is a subalgebra of the BMS${}_d$ algebra -- does not have unitary irreducible representation of the GK dimension $d-2$ with non-trivially realised translations.
This statement is a simple corollary of the Wigner classification of unitary irreducible representations of the Poincar\'{e} group, see \cite{Bekaert:2006py} for review.
The reason why one should expect that translations are realised non-trivially (more generally, to be not nilpotent) for the boundary fields is that, otherwise, the single-trace operators constructed out of them will also have trivially realised (nilpotent) translations, while this property is not valid for the on-shell bulk fields, which are expected to be dual to the single-trace operators on the boundary\footnote{For this reason, theories constructed e.g.  in \cite{Bagchi:2016bcd,Bagchi:2019xfx,Gupta:2020dtl,Rivera-Betancour:2022lkc,Baiguera:2022lsw} are unlikely to give suitable boundary duals to the typical Minkowski space theories. 
It is known for a long time that flat space contractions of singletons lead to trivially realised translations, see e.g. \cite{Flato:1978qz}.
Another proposal -- see  \cite{Bekaert:2022ipg} -- is to consider massless fields extended to representations of the complete BMS algebra as the dual theory candidates. These representations have translations realised non-trivially, though, as pointed out in \cite{Bekaert:2022ipg}, these are too large -- in the sense of having the GK dimension $d-1$ -- to fit into the familiar holographic framework.}. In other words, suitable flat space counterparts of the AdS/CFT boundary fields do not exist already at the level of representation theory. 
This means that to proceed with  flat space holography, we inevitably need to give up some of the basic features of the AdS/CFT correspondence\footnote{One way to avoid this conclusion is to suggest that flat space holography does not have any weak-weak regime, therefore, the representation theory arguments are inapplicable.}.

To understand which of the standard requirements might have to be relaxed, we will consider the setup of higher-spin holography. 
Higher-spin theories occupy an important place in the AdS/CFT duality. The simplest version of the AdS space higher-spin holography  \cite{Sezgin:2002rt,Klebanov:2002ja} relates higher-spin theories in the bulk with free conformal theories on the boundary. 
Due to the fact that the boundary theory is free, the AdS space higher-spin holography, essentially, reduces to a simple algebraic analysis based on the representation theory of the AdS space isometry algebra. Besides that, the free theory on the boundary has rich symmetry, which simultaneously plays the role of the global higher-spin algebra in the bulk. This symmetry is rich enough so that once it is known, the boundary theory can be identified \cite{Maldacena:2011jn,Boulanger:2013zza,Alba:2013yda}. An efficient approach to do that is to regard the higher-spin algebra as the universal enveloping algebra of the AdS space isometry algebra, quotiented out by generators, that annihilate the representation, carried by the on-shell field on the boundary, see \cite{Eastwood:2002su,Vasiliev:2003ev,Bekaert:2006zoe,Iazeolla:2008ix,Bekaert:2008sa,Boulanger:2011se,Joung:2014qya,Campoleoni:2021blr}. 

In the present paper, to address the issue of flat space higher-spin holography we use chiral higher-spin theories in flat space as a starting point\footnote{Chiral higher-spin theories and related structures were recently discussed in the context of celestial holography in \cite{Ren:2022sws,Monteiro:2022lwm,Bu:2022iak,Guevara:2022qnm}.}. Chiral higher-spin theories  were suggested in \cite{Ponomarev:2016lrm} based on the earlier analysis of \cite{Metsaev:1991mt,Metsaev:1991nb}. These theories were actively studied recently and, in particular, the associated chiral higher-spin algebra was found \cite{Ponomarev:2017nrr,Krasnov:2021nsq,Krasnov:2021cva,Skvortsov:2022syz,Sharapov:2022faa,Ponomarev:2022atv}. Having the higher-spin algebra at our disposal, we 
follow the steps mimicking the AdS space analysis to identify the dual theory. Namely, we  show that the flat space chiral higher-spin algebra can be constructed as the universal enveloping algebra of a certain chiral deformation of the Poincar\'{e} algebra, which is quotiented  by the ideal generated by a set of quadratic relations. These quadratic relations allow us to identify a representation of the chiral Poincar\'{e} algebra, which we regard as the flat space singleton.

This paper is organised as follows. We start by reviewing how the enveloping algebra construction works for higher-spin holography in the AdS space in section \ref{sec:2}. Next, in section \ref{sec:3} we first recall how the flat space chiral higher-spin algebra is defined and then show how the enveloping algebra approach can be extended to the flat space case. This allows us to define the flat space singleton representations, which we realise in two different ways in section \ref{sec:4}. Finally, in section \ref{sec:5} we give our conclusions.

\section{The AdS space case}
\label{sec:2}

In the present section we review the universal enveloping algebra approach towards the  higher-spin algebra in the AdS space. For earlier works, which include reviews and extensions, see \cite{Eastwood:2002su,Vasiliev:2003ev,Bekaert:2006zoe,Iazeolla:2008ix,Bekaert:2008sa,Boulanger:2011se,Joung:2014qya,Campoleoni:2021blr}.

We start by recalling how this construction works in general dimensions. Let the AdS${}_d$ space isometry algebra $g\equiv so(d-1,2)$ be defined as
\begin{equation}
\label{16sep1}
i[J_{MN},J_{RS}]=\tilde\eta_{NR}J_{MS}-\tilde\eta_{MS}J_{NR} -\tilde\eta_{SM}J_{RN}+\tilde\eta_{SN}J_{RM},
\end{equation}
where capital Latin letters take values from 0 to $d$ and $\tilde \eta = {\rm diag}(-,+,\dots,+,-)$. The universal enveloping algebra $U(g)$ is generated by polynomials in $J_{MN}$ modulo relations (\ref{16sep1}). According to the Poincar\'{e}-Birkhoff-Witt theorem the basis elements of  $U(g)$ can be chosen as symmetric products of $J_{MN}$.

The AdS space higher-spin algebra can be defined as $U(g)$ quotiented by the two-sided ideal
\begin{equation}
\label{16sep2}
hs_d\equiv U(g)/\langle I \rangle, \qquad  \langle I \rangle \equiv U(g)\; I\; U(g),
\end{equation}
where $I$ is generated by 
\begin{equation}
\label{16sep3}
I \equiv  \Big(J_{[MN} \odot J_{RS]} \Big) \oplus \Big( J^M{}_{(N} \odot J_{R)M} -\frac{2}{d+1} \tilde\eta_{NR}C_2\Big),
\end{equation}
\begin{equation}
\label{16sep4}
J_{MN} \odot J_{CD} \equiv \frac{1}{2} \{J_{MN},J_{CD} \}, \qquad C_2 \equiv \frac{1}{2}J_{MN}\odot J^{MN}.
\end{equation}
The role of the first term in (\ref{16sep3}) is to factor out  all products of $J_{MN}$ in $U(g)$, which have symmetries of Young diagrams with more than two rows, while the role of the second term is to factor out all traces of products of $J_{MN}$ in $U(g)$. Besides that, one can show that the vanishing of  (\ref{16sep3}) fixes the quadratic Casimir operator of $g$
\begin{equation}
\label{16sep5}
C_2\sim -\frac{(d+1)(d-3)}{4}.
\end{equation}
As a result, the higher-spin algebra is generated by traceless tensor representations of $so(d-1,2)$ with symmetries characterised by rectangular two-row Young diagrams.

Quadratic relations (\ref{16sep3}), (\ref{16sep5}) vanish when evaluated on the singleton representation of $so(d-1,2)$, that is the representation carried by the boundary field theory\footnote{In the given case ''the singleton'' just refers to the free massless scalar field on the boundary.}. Thus, the higher-spin algebra can be equivalently understood as the universal enveloping algebra $U(g)$ evaluated on the singleton representation. This property can be used to identify the singleton representation once the higher-spin algebra is known. To this end one just needs to consider quadratic elements of $U(g)$ and see which of them vanish for the higher-spin algebra. Vanishing of these quadratic generators uniquely defines the singleton representation.

In the present paper we will focus on the higher-spin algebras in the four-dimensional space-time. In this case it is convenient to make the Lorentz symmetry manifest by utilising $sl(2,\mathbb{C})$ spinors. Our conventions on $sl(2,\mathbb{C})$ spinors are given in appendix \ref{app:a}. To rewrite the previous discussion in these terms, one first needs to decompose $so(3,2)$ tensors into the Lorentz ones. In particular, the $so(3,2)$ generators are now decomposed into Lorentz transformations and deformed translations
\begin{equation}
\label{16sep6}
J_{MN} \quad \to \quad \{J_{mn}, P_m\equiv J_{md} \},
\end{equation}
where the Lorentz indices $m$, $n$ run from $0$ to $3$. Next, employing the vector-spinor dictionary, all Lorentz indices should be converted to spinor ones. For the generators of $so(3,2)$ this leads to
\begin{equation}
\label{16sep6x1}
\begin{split}
P_{\alpha\dot\alpha} \equiv (\sigma^m)_{\alpha\dot\alpha}P_m,\\
J_{\alpha\beta}\equiv J_{\beta\alpha}\equiv \frac{1}{2}(\sigma_m)_{\alpha}{}^{\dot\gamma}(\sigma_n)_{\beta\dot\gamma}J^{mn},\\
\bar J_{\dot\alpha\dot \beta}=\bar J_{\dot\beta\dot\alpha}=\frac{1}{2}(\sigma_m)^\gamma{}_{\dot\alpha}(\sigma_n)_{\gamma\dot\beta}J^{mn}.
\end{split}
\end{equation}

In these terms the quadratic relations (\ref{16sep3}) acquire the form
\begin{equation}
\label{16sep7}
\begin{split}
J^{\alpha\alpha}J_{\alpha\alpha} - \bar J^{\dot\alpha\dot\alpha}\bar J_{\dot\alpha\dot\alpha} \sim 0, \\
 P_{\beta\dot\alpha}J^{\beta}{}_\alpha + J^{\beta}{}_\alpha P_{\beta\dot\alpha}  -\text{c.c.} \sim 0,\\
 2P_{\alpha\dot\alpha}P_{\alpha\dot\alpha}- J_{\alpha\alpha}\bar J_{\dot\alpha\dot\alpha}- \bar J_{\dot\alpha\dot\alpha} J_{\alpha\alpha} \sim 0,\\
 P_{\beta\dot\alpha}J^{\beta}{}_\alpha + J^{\beta}{}_\alpha P_{\beta\dot\alpha}  +\text{c.c.}\sim 0,\\
J^{\alpha\alpha}J_{\alpha\alpha} + \bar J^{\dot\alpha\dot\alpha}\bar J_{\dot\alpha\dot\alpha} - 3 P^{\alpha\dot\alpha}P_{\alpha\dot\alpha}\sim 0,
\end{split}
\end{equation}
where c.c. refers to the complex conjugate expression, 
while (\ref{16sep5}) becomes
\begin{equation}
\label{16sep8}
C_2 = \frac{1}{4}\bar J^{\dot\alpha \dot\alpha}\bar J_{\dot\alpha\dot\alpha} + \frac{1}{4}J^{\alpha\alpha}J_{\alpha\alpha} + \frac{1}{2}P^{\alpha\dot\alpha}P_{\alpha\dot\alpha}\sim -\frac{5}{4}.
\end{equation}
For our conventions on symmetrisation and the use of repeated indices, see Appendix \ref{app:a}.

As it is not hard to see, with the ideal generated by (\ref{16sep7}), (\ref{16sep8}) quotiented out, one is left with the generators of the form $T_{\alpha(m),\dot\alpha(n)}$ with the total number of indices even. This spectrum matches the one that we discussed above using the $so(3,2)$ covariant terms once the latter is first rephrased in terms of Lorentz tensors and then converted to spinor notations\footnote{The necessary details can be found, e.g. in \cite{Didenko:2014dwa}.}.

The AdS${}_4$ higher-spin algebra \cite{Fradkin:1986ka,Vasiliev:1999ba} can be realised explicitly in terms of the associative star product
\begin{equation}
\label{16sep9}
(f\star g)(\lambda,\bar\lambda) \equiv f(\lambda,\bar\lambda) {\rm exp}\left[i \epsilon^{\dot\alpha\dot\beta}\frac{\overleftarrow\partial}{\partial \bar\lambda^{\dot\alpha}} \frac{\overrightarrow\partial}{\partial \bar\lambda^{\dot\beta}}+
i \epsilon^{\alpha\beta}\frac{\overleftarrow\partial}{\partial \lambda^{\alpha}} \frac{\overrightarrow\partial}{\partial \lambda^{\beta}}
 \right] g(\lambda,\bar\lambda). 
\end{equation}
The higher-spin algebra generators are given by polynomials $f(\lambda,\bar\lambda)$ with the  total degree even, while the Lie algebra bracket is just the star product commutator
\begin{equation}
\label{16sep9x1}
[f,g](\lambda,\bar\lambda)\equiv (f\star g)(\lambda,\bar\lambda)-(g\star f)(\lambda,\bar\lambda).
\end{equation}

To make contact with the previous discussion, we first note that  quadratic monomials 
\begin{equation}
\label{16sep10}
\begin{split}
\bar J_{\dot\alpha\dot\alpha}\equiv \frac{1}{2}\bar\lambda_{\dot \alpha}\bar\lambda_{\dot\alpha}, \\
P_{\alpha\dot\alpha}\equiv\frac{1}{2}\lambda_\alpha\bar\lambda_{\dot\alpha},\\
J_{\alpha\alpha}\equiv \frac{1}{2}\lambda_\alpha\lambda_{\alpha}
\end{split}
\end{equation}
generate the $so(3,2)$ subalgebra of the higher-spin algebra \cite{Fradkin:1986ka,Vasiliev:1999ba}. It is then straightforward  to check that (\ref{16sep7}), (\ref{16sep8}) are identically satisfied when the product in (\ref{16sep7}), (\ref{16sep8}) is understood as the star product. In other words, the star product realisation of the higher-spin algebra in AdS${}_4$ automatically resolves the ideal, which annihilates the singleton representation.

\section{Flat space case}
\label{sec:3}

In this section we will show how the discussion of the previous section can be carried over to the flat space case. However, in contrast to the previous section, here we will start from the known realisation of the flat space higher-spin algebra, which will then be put into the universal enveloping algebra framework. This will allow us to find the quadratic relations that define the flat space counterpart of the singleton representation.

To start, we remind that the higher-spin algebra associated with chiral higher-spin theories in flat space \cite{Metsaev:1991mt,Metsaev:1991nb,Ponomarev:2016lrm} can be formulated in terms of the product \cite{Ponomarev:2017nrr,Krasnov:2021nsq,Krasnov:2021cva,Skvortsov:2022syz,Sharapov:2022faa,Ponomarev:2022atv}
\begin{equation}
\label{19sep1}
(f\;\bar\circ\; g)(\lambda,\bar\lambda) \equiv f(\lambda,\bar\lambda) {\rm exp}\left[i \epsilon^{\dot\alpha\dot\beta}\frac{\overleftarrow\partial}{\partial \bar\lambda^{\dot\alpha}} \frac{\overrightarrow\partial}{\partial \bar\lambda^{\dot\beta}} \right] g(\lambda,\bar\lambda). 
\end{equation}
This product is associative and can be understood as a contraction of the star product (\ref{16sep9}), in which the $\lambda$ spinors become commutative. Equivalently, (\ref{19sep1}) is the star product on the $\bar\lambda$ variables, with the coefficients valued in functions of $\lambda$. Cleary, if two such functions are supported at different $\lambda$'s, their product  vanishes.
On the contrary, if $f$ and $g$ are supported at the same $\lambda$ their $\bar\circ$-product reduces to the usual star product in $\bar\lambda$. Thus, intuitively, the associative algebra defined by (\ref{19sep1}) can be viewed as  the direct sum of infinitely many copies of the star product algebra on variable $\bar\lambda$ with different copies labelled by $\lambda$. A similar structure is characteristic of flat space singleton representations, to be discussed later. 

Similarly to the AdS space case, we consider quadratic monomials
\begin{equation}
\label{19sep2}
\begin{split}
\bar J_{\dot\alpha\dot\alpha}\equiv \frac{1}{2}\bar\lambda_{\dot \alpha}\bar\lambda_{\dot\alpha}, \\
P_{\alpha\dot\alpha}\equiv\frac{1}{2}\lambda_\alpha\bar\lambda_{\dot\alpha},\\
L_{\alpha\alpha}\equiv \frac{1}{2}\lambda_\alpha\lambda_{\alpha}.
\end{split}
\end{equation}
Here $\bar J$ generates a chiral part of the Lorentz subalgebra $sl(2,\mathbb{C})$ with respect to which $P_{\alpha\dot\alpha}$ transforms according to its index structure. We will often refer to the $\bar J$ part of the Lorentz algebra as the antiholomorphic Lorentz algebra.
The commutator of $P$'s is given by 
\begin{equation}
\label{19sep3}
[P_{\alpha\dot\alpha},P_{\beta\dot\beta}]=-i\epsilon_{\dot\alpha\dot\beta}L_{\alpha\beta},
\end{equation}
while $L$ commutes with itself and with other generators. We will refer to the algebra generated by (\ref{19sep2}) as the {\em deformed chiral Poincar\'{e} algebra}.
For more details on the algebras we are dealing with in this paper see  appendix \ref{app:alg}.

The algebra generated by (\ref{19sep2}) is, obviously, not the Poincar\'{e} algebra. Firstly, unlike translations, $P$'s in (\ref{19sep2}) do not commute. Secondly, $L$ does not generate the missing part of $sl(2,\mathbb{C})$, instead, it provides a central extension of the algebra generated by $P$ and $\bar J$. This can be reconciled with the fact that we are dealing with massless higher-spin fields in the Minkowski space in the following way \cite{Ponomarev:2022atv}. Firstly, massless higher-spin fields appear to be in the representation for which $L=0$ and, therefore, $P$'s actually commute as translations should. Secondly, the theory still has the
missing part of the Lorentz algebra as  symmetry, though, this symmetry is not a part of the chiral higher-spin algebra. 

We now evaluate the flat space counterparts of the AdS quadratic relations (\ref{16sep7}), (\ref{16sep8}) by using the $\bar\circ$ product instead of the star product. After some straightforward algebra we find
\begin{equation}
\label{19sep4}
\begin{split}
P^{\alpha\dot\alpha}\;\bar \circ\; P_{\alpha\dot\alpha}= 0,\\
\bar J^{\dot\alpha\dot\alpha}\;\bar \circ\; \bar J_{\dot\alpha\dot\alpha}=-\frac{3}{2},\\
P^{\dot\beta}{}_{\alpha} \;\bar \circ\; \bar J_{\dot\beta\dot\alpha}+\bar J_{\dot\beta\dot\alpha} \;\bar \circ\; P^{\dot\beta}{}_{\alpha}=0,\\
L^{\alpha\alpha} \;\bar \circ\; L_{\alpha\alpha}=0,\\
P^{\alpha}{}_{\dot\beta} \;\bar \circ\; L_{\alpha \beta}+ L_{\alpha \beta} \;\bar \circ\; P^{\alpha}{}_{\dot\beta}=0,\\
 2P_{\alpha\dot\alpha} \;\bar \circ\; P_{\alpha\dot\alpha}- L_{\alpha\alpha}  \;\bar \circ\; \bar J_{\dot\alpha\dot\alpha}- \bar J_{\dot\alpha\dot\alpha}  \;\bar \circ\; L_{\alpha\alpha} = 0.
\end{split}
\end{equation}
These relations specify the flat space singleton representations, which will be explicitly realised in the following section.

\section{Realisations of the flat space singleton representation}
\label{sec:4}

Considering that $L$ commute, it is convenient to choose a basis in the representation space
as eigenstates of $L$.
 Moreover, for the flat space singleton representation one has (\ref{19sep4})
 \begin{equation}
 \label{24sep1}
 L_{\alpha\alpha}L^{\alpha\alpha}=0 \qquad \Leftrightarrow \qquad {\rm det}(L)=0.
 \end{equation}
Considering that $L_{\alpha\alpha}$ is also symmetric, it can be presented as\footnote{Note that despite the last line of (\ref{19sep2}) looks identical to (\ref{19dep5}), their meaning is different: while in (\ref{19dep5}) $L$ is the action of a Lie algebra generator in the flat space singleton representation, in (\ref{19sep2}) it represents the action of the associative algebra generator $L$ in the higher-spin algebra representation. In particular, as a Lie algebra generator $L$ acts trivially in the higher-spin algebra representation.}
\begin{equation}
\label{19dep5}
L_{\alpha\alpha}\equiv \frac{1}{2}\lambda_\alpha\lambda_{\alpha}
\end{equation}
with $\lambda$ being the usual commuting variables. Since $L$ commute with all generators of the deformed chiral Poincar\'{e} algebra, they serve as representation labels, similar to $m^2$ for the Poincar\'{e} algebra. Alternatively, representations can be labelled with $\lambda$. With the latter labelling representations associated with $\lambda$ and with $-\lambda$ are equivalent. 

\subsection{Manifest $\bar J$ symmetry}
\label{sec:41}

We will start by realising the flat space singleton representation labelled with $\lambda$ in the form, which makes the $\bar J$ symmetry manifest due to the use of $\bar\lambda$ spinors. This means that a representation will be carried by a function $\varphi_\lambda(\bar\lambda)$, on which $\bar J$ act as
\begin{equation}
\label{19sep6}
\begin{split}
\bar J_{\dot\alpha\dot\alpha}\varphi_\lambda(\bar\lambda)=2 i \bar\lambda_{\dot\alpha}\frac{\partial \varphi_\lambda(\bar\lambda)}{\partial \bar\lambda^{\dot\alpha}}.
\end{split}
\end{equation}
Such a realisation can be considered as the flat space version of the approach to the AdS space singletons, which makes the $sl(2,\mathbb{C})$ symmetry manifest \cite{Dirac:1963ta,Ponomarev:2021xdq}.

In these terms, the second equation in (\ref{19sep4}) gives\footnote{We would like to remark that (\ref{19sep7}) can be regarded as a candidate equation of motion for $\varphi_\lambda(\bar\lambda)$. When converted back to the vector notations, $\varphi_\lambda(\bar\lambda)$ becomes a function on the light-cone times $U(1)$, with the $U(1)$ factor  being responsible for carrying the spin labels. Considering that the light-cone is isomorphic to the Minkowski space boundary, (\ref{19sep7}) has some features of a boundary theory equation of motion. It would be interesting to explore this in future further. In particular, it would be interesting to extend this construction off shell in a way that the deformed chiral Poincar\'{e} algebra is a symmetry of the action.}
\begin{equation}
\label{19sep7}
2 (\bar N +2)\bar N  \varphi_\lambda(\bar\lambda) = -\frac{3}{2}\varphi_\lambda(\bar\lambda),
\end{equation}
where
\begin{equation}
\label{19sep8}
\bar N\equiv \bar\lambda^{\dot\alpha}\frac{\partial}{\partial\bar\lambda^{\dot\alpha}}.
\end{equation}
Equation (\ref{19sep7}) has two solutions 
\begin{equation}
\label{19sep9}
\bar N =-\frac{1}{2}, \qquad \bar N =-\frac{3}{2},
\end{equation}
accordingly, $\varphi_\lambda$ splits into two subspaces with fixed homogeneity degrees in $\bar \lambda$
\begin{equation}
\label{19sep10}
\bar N \varphi_\lambda^{-\frac{1}{2}}= -\frac{1}{2} \varphi_\lambda^{-\frac{1}{2}}, \qquad \bar N \varphi_\lambda^{-\frac{3}{2}}= -\frac{3}{2} \varphi_\lambda^{-\frac{3}{2}}.
\end{equation}

Then $\bar J$ covariance requires that deformed translations are representated as
\begin{equation}
\label{19sep11}
P_{\alpha\dot\alpha}\varphi_\lambda^{-\frac{3}{2}}=A\lambda_{\alpha}\bar\lambda_{\dot\alpha}\varphi_\lambda^{-\frac{3}{2}}, \qquad
P_{\alpha\dot\alpha}\varphi_\lambda^{-\frac{1}{2}}=B\lambda_{\alpha}\frac{\partial}{\partial\bar\lambda^{\dot\alpha}}\varphi_\lambda^{-\frac{1}{2}}.
\end{equation}
The unknown coefficients $A$ and $B$ are then fixed from the requirement that (\ref{19sep3}) holds. This leads to 
\begin{equation}
\label{19sep12}
AB=i.
\end{equation}
This constraint allows us to solve for $B$ in terms of $A$. The value of $A$, in turn, is inessential, as it can be fixed to any given one by relative rescalings of $\varphi^{-\frac{1}{2}}$ and $\varphi^{-\frac{3}{2}}$. 

It can be checked that for representation (\ref{19sep6}), (\ref{19sep11}), (\ref{19sep12}) all conditions (\ref{19sep4}) are satisfied. Finally, we note that this representation is not irreducible: it can be split into two subspaces analogously to the splitting of the AdS singleton representation into positive and negative energy modes, see appendix \ref{app:b}. 

\subsection{Single-variable realisation}
\label{sec:42}

In this section we will consider a single-variable realisation of the flat space singleton representation. It is somewhat similar to the oscillator realisation in the AdS${}_4$ case first proposed in \cite{Dirac:1963ta}.

In this approach, the representation space is realised by a function $c_\lambda (a)$  of a single variable $a$.
The anti-holomorphic Lorentz  commutation relations 
\begin{equation}
\label{24aug1}
[\bar J_{\dot 1\dot 1},\bar J_{\dot 2\dot 2}]=4i\bar J_{\dot 1\dot 2}, \qquad [\bar J_{\dot 1\dot 1},\bar J_{\dot 1\dot 2}]=2i 
\bar J_{\dot 1\dot 1},\qquad 
[\bar J_{\dot 2\dot 2},\bar J_{\dot 1\dot 2}]=-2i 
\bar J_{\dot 2\dot 2}
\end{equation}
can be realised with 
\begin{equation}
\label{24aug2}
\bar J_{\dot 1\dot 1}= X a^2, \qquad \bar J_{\dot 2\dot 2}=Y \frac{\partial^2}{\partial a^2}, \qquad
\bar J_{\dot 1\dot 2}= iXY (a\frac{\partial}{\partial a}+\frac{1}{2}),
\end{equation}
where
\begin{equation}
\label{24aug3}
XY=-1.
\end{equation}
Here $X$ can be fixed to any convenient value, which is related to the freedom to rescale the $a$ variable, while $Y$ is then defined in terms of $X$ via (\ref{24aug3}).

The components of an auxiliary dotted spinor $\bar\pi_{\dot\alpha}$ can be realised as 
\begin{equation}
\label{24aug4}
\bar \pi_{\dot 1}=W a, \qquad \bar \pi_{\dot 2}=Z \frac{\partial}{\partial a},
\end{equation}
where
\begin{equation}
\label{24aug5}
YW = -iZ.
\end{equation}
Evaluating the commutator of $\bar \pi_{\dot \alpha}$ we find
\begin{equation}
\label{24aug6}
[\bar \pi_{\dot \alpha},\bar\pi_{\dot \beta}]=WZ \varepsilon_{\dot\alpha\dot\beta}.
\end{equation}
It is then natural to look for the deformed translations in the form
\begin{equation}
\label{24aug7}
P_{\alpha\dot\alpha}=\lambda_\alpha \bar\pi_{\dot\alpha}.
\end{equation}
To reproduce the correct commutation relations (\ref{19sep3}), one has to require
\begin{equation}
\label{24aug8}
WZ=-\frac{i}{2}.
\end{equation}
Once $X$ and $Y$ are fixed, (\ref{24aug5}) and (\ref{24aug8}) allow us to fix $W$ and $Z$ up to an irrelevant sign, which can be absorbed into $\lambda\to -\lambda$. It can be checked that the representation we have just defined satisfies (\ref{19sep4}).

To summarise, (\ref{24aug2}), (\ref{24aug7}), (\ref{24aug4}) together with constraints (\ref{24aug3}), (\ref{24aug5}) and (\ref{24aug8}) provide the oscillator-like realisation of the flat space singleton representation. This representation is irreducible for given $\lambda$.

\subsection{Remarks}

In this section we would like to make a couple of remarks concerning the flat space singleton representations we constructed above.

To start, we note that in the split signature $(+,+,-,-)$ the single-variable representation of section \ref{sec:42}  is unitary. Indeed, with a choice 
\begin{equation}
\label{21sep1}
X=1, \qquad Y=-1, \qquad Z=\frac{i}{\sqrt{2}}, \qquad W=-\frac{1}{\sqrt{2}}
\end{equation}
operators (\ref{24aug2}), (\ref{24aug7}) are self-adjoint with respect to the inner product
\begin{equation}
\label{21sep2}
(b_\lambda,c_\lambda)\equiv \int da\; \bar b_\lambda(a)c_\lambda(a)
\end{equation}
in the sense that for all of them
\begin{equation}
\label{17dec1}
(Ob_\lambda,c_\lambda) = (b_\lambda, O c_\lambda).
\end{equation}
Above  the bar denotes the complex conjugation and $\lambda$ is understood to be real. Thus, despite the construction of section \ref{sec:42} has some features of the oscillatorial realisation of singletons in AdS, the inner product (\ref{21sep2}) is different from the natural oscillatorial one. 

As a second remark, we note that the flat space singleton representation can be naturally promoted to a representations of the algebra, which, in addition, features $J$ generators of the Lorentz algebra. In appendix \ref{app:alg} it is referred to as the {\em{$L$-extended Poincar\'{e} algebra}}. To this end, one just needs to assume that the flat space singleton wave function also depends on $\lambda$ and $J$ acts in the standard way
\begin{equation}
\label{7oct4}
 J_{\alpha \alpha}=2 i \lambda_{\alpha}\frac{\partial }{\partial \lambda^{\alpha}}.
\end{equation}
Then, $c_\lambda(a)$ with different $\lambda$'s are mixed by $J$ transformations. Still,  one can impose two invariant conditions on the wave function
\begin{equation}
\label{7oct5}
c_{-\lambda}(-a)=\pm c_\lambda(a).
\end{equation}
In other words, the representation space $c_\lambda(a)$ can be split into invariant subspaces of even and odd total homogeneity degrees in $a$ and $\lambda$.
By analogy with the AdS case, we will refer to the even representation as the scalar flat space singleton, while the odd one will be referred to as the spinor flat space singleton.

Finally, we remark that the flat space singleton representation can be naturally promoted to a parity-invariant representation of the {\em $(L,\bar L)$-extended Poincar\'{e} algebra}, which extends the $L$-extended Poincar\'{e} algebra in a parity-invariant way, see appendix \ref{app:alg} for the precise definition. To this end, it is convenient to start from the flat space singleton representation realised  as in section \ref{sec:41}. In addition to $\varphi_\lambda(\bar\lambda)$, one also considers its complex conjugate
$\bar\varphi_{\bar\lambda}(\lambda)$, which satisfies
\begin{equation}
\label{7oct6}
2 ( N +2) N  \bar\varphi_{\bar \lambda}(\lambda) = -\frac{3}{2}\bar\varphi_{\bar\lambda}(\lambda),
\end{equation}
where
\begin{equation}
\label{7oct7}
 N\equiv \lambda^{\alpha}\frac{\partial}{\partial\lambda^{\alpha}}.
\end{equation}
The Lorentz generators act on $\bar\varphi$ in the standard way (\ref{19sep6}), (\ref{7oct4}), while the deformed translations act by
\begin{equation}
\label{7oct8}
P_{\alpha\dot\alpha}\bar\varphi_{\bar\lambda}^{-\frac{3}{2}}=\bar A\lambda_{\alpha}\bar\lambda_{\dot\alpha}\bar\varphi_{\bar\lambda}^{-\frac{3}{2}}, \qquad
P_{\alpha\dot\alpha}\bar\varphi_{\bar\lambda}^{-\frac{1}{2}}=\bar B\bar\lambda_{\dot\alpha}\frac{\partial}{\partial\lambda^{\alpha}}\bar\varphi_{\bar\lambda}^{-\frac{1}{2}}.
\end{equation}
Then, one has 
\begin{equation}
\label{7oct9}
 L_{\alpha\alpha} \bar\varphi =0, \qquad \bar L_{\dot\alpha\dot\alpha} \bar \varphi = \frac{1}{2}\bar\lambda_{\dot\alpha}\bar\lambda_{\dot\alpha}\bar \varphi.
\end{equation}

\section{Conclusions}
\label{sec:5}

In the present paper we studied the flat space chiral higher-spin algebra focusing on its universal enveloping algebra realisation. We found that it can, indeed, be constructed as the universal enveloping algebra of the deformed chiral Poincar\'{e} algebra, quotiented out by a set of quadratic relations. These quadratic relations allow us to identify the flat space counterpart of the $so(3,2)$ singleton representation. This provides us with a concrete candidate for a theory, which is holographically dual to chiral higher-spin theories in flat space or their extensions discussed in \cite{Ponomarev:2022atv}. 

This construction has a number of features, which make it different from how holography works in the AdS space and from what is usually discussed in the context of  flat space holography. As we mentioned in the introduction, basic representation theory arguments imply that some changes in the holographic setup with the Minkowski space as the bulk are inevitably required. Let us highlight the important peculiarities of our construction.

The key difference with other approaches to flat holography is that in our case the symmetry algebra of the dual theory is neither given by the BMS algebra nor by the Poincar\'{e} algebra, instead, it is  a certain chiral extension of the Poincar\'{e} algebra, for which translations do not commute. 
As discussed in appendix \ref{app:alg},  this extended algebra can be connected to the so-called Maxwell algebra \cite{Bacry:1970ye,Bacry:1970du,Schrader:1972zd,Bonanos:2008ez,Gomis:2017cmt}.
The commutator of translations is given by (\ref{19sep3}), where $L\sim\lambda\lambda$ commutes with all generators and, hence, serves as a representation label. The massless higher-spin fields in the bulk have $L$ vanishing, which explains how one can recover the usual Poincar\'{e} algebra in the bulk. Instead, for the dual flat space singleton $L$ can be non-zero, which impedes its interpretation as the boundary theory with the boundary understood in the usual geometric sense. 

An important difference between the AdS space higher-spin holography and the setup considered here is that the flat space singleton representation is not unique, instead, these are labelled by a continuous label -- the $sl(2,\mathbb{C})$ spinor $\lambda$. This property can be traced to the properties of the flat space higher-spin algebra, which has an analogous structure. The higher-spin algebra is still the symmetry of the flat space singleton representation, though, elements of the algebra act non-trivially on the flat space singleton representation only if both are associated with the same value of $\lambda$.

The obvious next step is to explore whether the flat space singleton representation discussed here features any generalisation of the Flato-Fronsdal theorem \cite{Flato:1978qz}. The basic counting of degrees of freedom suggests that such a theorem may exist. Indeed, at fixed $\lambda$, massless higher-spin fields are functions of the two-component spinor $\bar\lambda$. In turn, at fixed $\lambda$ the representation space of the flat space singleton is given by a function  $c_{\lambda}(a)$ of a single variable $a$. Thus, the tensor product of two flat space singletons at fixed $\lambda$ has the right number of degrees of freedom to match the states of the tower of massless higher-spin fields. This is how the representation theory obstacle, that we outlined in the introduction, can be resolved.  Moreover,  a tensor product at fixed $\lambda$ seems natural, considering that the higher-spin amplitudes it should eventually give rise to holographically are non-trivial only when all fields on the external lines have equal $\lambda$ components \cite{Ponomarev:2022atv}. In fact, if the flat space counterpart of the Flato-Fronsdal theorem does exist, the correlators in a theory of flat space singletons should  be inevitably consistent with the results on the higher-spin side due to the constraining role of the higher-spin symmetry.

\acknowledgments

We would like to thank M. Grigoriev,  V. Didenko and E. Skvortsov  for fruitful discussions on various subjects related to the paper.
I would also like to thank the participants of ''Higher Spin Gravity and its Applications'' workshop held at Asia Pacific Center for Theoretical Physics for interesting comments as well as the organisers for their hospitality. 
 This work was supported by Russian Science Foundation Grant 18-72-10123.

\appendix

\section{Conventions}
\label{app:a}

In this appendix we collect our conventions on  $sl(2,\mathbb{C})$ spinors as well as some useful formulae.

A Lorentz vector can be converted to a bispinor via 
\begin{equation}
\label{5nov4}
v_{\alpha\dot\alpha}\equiv v_a (\sigma^a)_{\alpha\dot\alpha},
\end{equation}
where 
\begin{equation}
\label{5nov3}
\sigma^0 = 
\left(\begin{array}{cccc}
1 &&& 0\\
0 && &1
\end{array}\right), \quad \sigma^1 = 
\left(\begin{array}{cccc}
0 & && 1\\
1& && 0
\end{array}\right), 
\quad 
 \sigma^2 = 
\left(\begin{array}{ccc}
0 && -i\\
i&& 0
\end{array}\right), \quad
 \sigma^3 = 
\left(\begin{array}{ccc}
1 && 0\\
0&& -1
\end{array}\right)
\end{equation}
are the Pauli matrices. Spinor indices are raised and lowered using the following convention 
 \begin{equation}
\label{5nov6}
\lambda^\alpha = \epsilon^{\alpha\beta} \lambda_\beta, \qquad \lambda_\beta=\epsilon_{\beta\gamma} \lambda^\gamma,
\end{equation}
where
\begin{equation}
\label{5nov7}
\epsilon^{\alpha\beta}=\epsilon^{\dot\alpha\dot\beta}=
\left(
\begin{array}{cccc}
0&&& 1\\
-1&&&0
\end{array}
\right) = -\epsilon_{\alpha\beta}=-\epsilon_{\dot\alpha\dot\beta}.
\end{equation}

Relation (\ref{5nov4}) can be inverted as follows
 \begin{equation}
 \label{5nov8}
 v_a=-\frac{1}{2}(\sigma_a)^{\dot\alpha\alpha}v_{\alpha\dot\alpha}.
 \end{equation}
 To this end one needs to use
 \begin{equation}
 \label{5nov9}
 (\sigma^a)_{\alpha\dot\alpha}(\sigma_a)_{\beta\dot\beta}=-2\epsilon_{\alpha\beta}\epsilon_{\dot\alpha\dot\beta},
 \qquad
  (\sigma^a)^{\alpha\dot\alpha}(\sigma_a)^{\beta\dot\beta}=-2\epsilon^{\alpha\beta}\epsilon^{\dot\alpha\dot\beta}.
 \end{equation}
Formulae (\ref{5nov4}), (\ref{5nov8}) can be used to convert (deformed) translation between the spinor and vector representations.

This machinery can be straightforwardly extended to include tensors of arbitrary symmetry. 
In particular, for antisymmetric rank two tensor $J_{ab}=-J_{ba}$ one has 
 \begin{equation}
 \label{5nov10}
 J_{\alpha\dot\alpha,\beta\dot\beta}\equiv J^{ab}(\sigma_a)_{\alpha\dot\alpha}(\sigma_b)_{\beta\dot\beta}.
 \end{equation}
It is easy to show that antisymmetry of $J^{ab}$ implies that $J_{\alpha\dot\alpha,\beta\dot\beta}$ is of the form
  \begin{equation}
 \label{5nov11}
 J_{\alpha\dot\alpha,\beta\dot\beta} = \epsilon_{\alpha\beta}\bar J_{\dot\alpha\dot\beta}+
 \epsilon_{\dot\alpha\dot\beta}J_{\alpha\beta},
 \end{equation}
 with $J_{\alpha\beta}$ and $\bar J_{\dot\alpha\dot\beta}$ symmetric.
  Here
 \begin{equation}
 \label{5nov12}
 \begin{split}
J_{\alpha\beta}=J_{\beta\alpha}\equiv \frac{1}{2}(\sigma_a)_\alpha{}^{\dot\gamma} (\sigma_b)_{\beta\dot\gamma}J^{ab},\\
\bar J_{\dot\alpha\dot\beta}=\bar J_{\dot\beta\dot\alpha} = \frac{1}{2}(\sigma_a)^\gamma{}_{\dot\alpha} (\sigma_b)_{\gamma\dot\beta}J^{ab}.
\end{split}
 \end{equation}
For real $J_{ab}$ bispinors  $J_{\alpha\beta}$ and $\bar J_{\dot\alpha\dot\beta}$ are complex conjugate to each other. One can invert (\ref{5nov12}), which leads to
 \begin{equation}
\label{5nov13}
J_{ab} = \frac{1}{4} (\sigma_a)^{\dot\alpha\alpha}(\sigma_b)^{\dot\beta \beta}(\epsilon_{\alpha\beta}\bar J_{\dot\alpha\dot\beta}+
\epsilon_{\dot\alpha\dot\beta}J_{\alpha\beta}).
\end{equation}
These relations can be used to convert Lorentz generators from the vector representation to the spinor one and back. 

We also use a convention according to which indices of a symmetric tensor are denoted by the same letter. In particular,  symmetric tensors in (\ref{5nov12}) are often denoted $J_{\alpha\alpha}$ and $\bar J_{\dot\alpha\dot\alpha}$. Moreover, the same convention is used to indicate that a tensor expression should be projected into the symmetric part. For instance, 
\begin{equation}
\label{13dec1}
P_{\alpha\dot\alpha}S_{\alpha\dot\alpha} = \frac{1}{4} \left(P_{\alpha_1\dot\alpha_1}S_{\alpha_2\dot\alpha_2}+P_{\alpha_1\dot\alpha_2}S_{\alpha_2\dot\alpha_1}+P_{\alpha_2\dot\alpha_1}S_{\alpha_1\dot\alpha_2}+P_{\alpha_2\dot\alpha_2}S_{\alpha_1\dot\alpha_1} \right).
\end{equation}
Finally, for tensors with multiple symmetric indices we write explicitly one of them an indicate the number of symmetric indices in brackets. For example, $T_{\alpha(m),\dot\alpha(n)}$ is symmetric in $m$ indices $\alpha$ as well as symmetric in $n$ indices $\dot\alpha$.

\section{Extended Poincar\'{e} algebras}
\label{app:alg}

The deformed chiral Poincar\'{e} algebra is generated by $P$, $\bar J$ and $L$ generators. The non-trivial commutation relations are
\begin{equation}
\label{7oct1}
\begin{split}
[\bar J_{\dot\alpha\dot\alpha},\bar J_{\dot\beta\dot\beta}]&=-4 i \epsilon_{\dot\alpha\dot\beta}\bar J_{\dot\alpha\dot \beta}, 
\qquad 
[ P_{\alpha\dot\alpha},\bar J_{\dot\beta\dot\beta}] = -2i \epsilon_{\dot\alpha\dot\beta}P_{\alpha\dot\beta},\\
[P_{\alpha\dot\alpha},P_{\beta\dot\beta}]&=-i\epsilon_{\dot\alpha\dot\beta}L_{\alpha\beta}.
\end{split}
\end{equation}

Along with this algebra, we also consider the $L$-extended Poincar\'{e} algebra. In addition to $P$, $\bar J$ and $L$, it also features Lorentz generators $J$. 
It is defined by non-trivial commutation relations  (\ref{7oct1}) together with 
\begin{equation}
\label{7oct2}
\begin{split}
[ J_{\alpha\alpha}, J_{\beta\beta}]&=-4 i \epsilon_{\alpha\beta} J_{\alpha \beta}, 
\qquad 
[ P_{\alpha\dot\alpha}, J_{\beta\beta}] = -2i \epsilon_{\alpha\beta}P_{\alpha\dot\beta},\\
[L_{\alpha\alpha},J_{\beta\beta}]&=-4i\epsilon_{\alpha\beta}L_{\alpha\beta}.
\end{split}
\end{equation}

Finally, it seems natural to consider a parity-invariant version of the $L$-extended algebra. It is generated by $P$, $J$, $\bar J$, $L$ and $\bar L$ generators. 
The commutator of translations, instead of (\ref{7oct1}) is given by
\begin{equation}
\label{7oct3}
[P_{\alpha\dot\alpha},P_{\beta\dot\beta}]=-i\epsilon_{\dot\alpha\dot\beta}L_{\alpha\beta}  -i\epsilon_{\alpha\beta}\bar L_{\dot\alpha\dot\beta}.
\end{equation}
The $\bar L$ generator transforms under $\bar J$ as
\begin{equation}
\label{7oct4x1}
\begin{split}
[\bar L_{\dot\alpha\dot\alpha},\bar J_{\dot\beta\dot\beta}]=-4i\epsilon_{\dot\alpha\dot\beta}\bar L_{\dot\alpha\dot\beta}
\end{split}
\end{equation}
and commutes with other generators. The remaining non-trivial commutation relations are as in (\ref{7oct1}), (\ref{7oct2}). We will refer to this algebra as the $(L,\bar L)$-extended Poincar\'{e} algebra. For $\bar L=(L)^*$ this algebra is parity-invariant. 

The $(L,\bar L)$-extended Poincar\'{e}  algebra that we encountered here is also known as the Maxwell algebra\footnote{We would like to thank X. Bekaert for pointing this to us.}.
It first appeared in \cite{Schrader:1972zd} and can be connected to particle motion in a constant electro-magnetic background, $\partial_c F_{ab}=0$, \cite{Bacry:1970ye,Schrader:1972zd,Bacry:1970du,Bonanos:2008ez}. For recent discussions and further references, see \cite{Gomis:2017cmt}. 
The $L$-extended Poincar\'{e}  algebra is then the Maxwell algebra with a self-dual field strength, while the deformed chiral Poincar\'{e} algebra is a chiral subalgebra of the former. In the context of our analysis, it is worthwhile exploring the existing literature for systematics of the associated representation theory, the homogeneous spaces these algebras naturally act on, etc.

\section{Splitting the flat space singleton representation}
\label{app:b}

In this appendix we show that representation (\ref{19sep6}), (\ref{19sep11}), (\ref{19sep12}) can be split into two invariant subspaces in complete analogy with the splitting of the AdS singleton representation into positive and negative energy modes \cite{Ponomarev:2021xdq}.

To start, we first solve the homogeneity degree constraints (\ref{19sep10}) in the form
\begin{equation}
\label{19sep13}
\phi(\bar z)=\varphi^{\bar N}(\bar\lambda^{\dot 1}, 1), \qquad \varphi^{\bar N}(\bar\lambda)=(\bar\lambda^{\dot 2})^{\bar N}\phi(\bar z), \qquad \bar z\equiv \frac{\bar\lambda^{\dot 1}}{\bar\lambda^{\dot 2}}.
\end{equation}
Splitting of the representation (\ref{19sep6}), (\ref{19sep11}), (\ref{19sep12}) is based on the fact that there is an $\bar J$ invariant intertwining operator between representations carried by $\varphi^{-\frac{1}{2}}$ and  $\varphi^{-\frac{3}{2}}$
\begin{equation}
\label{19sep14}
\phi^{-\frac{3}{2}}(\bar z)= O \phi^{-\frac{1}{2}}(\bar z)\equiv d\frac{i}{2}\int (\bar z-\bar z_1)^{-\frac{3}{2}} \phi^{-\frac{1}{2}}(\bar z_1)d\bar z_1
\end{equation}
where $d$ is an arbitrary normalisation factor. It has the form of the chiral part of the $sl(2,\mathbb{C})$ shadow transform and sometimes it is referred to as the light transform, see e.g. \cite{Strominger:2021lvk,Sharma:2021gcz} for discussions in the celestial holography context. 

The intertwining operator (\ref{19sep14}) by construction is $\bar J$ invariant. Below we will require it to be invariant with respect to the deformed translations, which will ensure invariance with respect to the complete chiral Poincar\'{e} algebra, as well as will fix $d$. To this end, we will focus on the action of $P_{1\dot 1}$ component
\begin{equation}
\label{19sep15}
\begin{split}
\tilde\phi^{-\frac{1}{2}}(\bar z)\equiv  P_{1\dot 1}\phi^{-\frac{3}2}(\bar z)=-A\lambda_1 \phi^{-\frac{3}{2}}(\bar z),\\
\tilde\phi^{-\frac{3}{2}}(\bar z)\equiv  P_{1\dot 1}\phi^{-\frac{1}2}(\bar z)=B\lambda_1 \frac{\partial\phi^{-\frac{1}{2}}(\bar z)}{\partial \bar z}.
\end{split}
\end{equation}
Invariance with respect to $P_{1\dot 1}$ amounts to the fact that once (\ref{19sep14}) is satisfied, then
\begin{equation}
\label{14sep9}
\tilde\phi^{-\frac{3}{2}}(\bar z)= O \tilde\phi^{-\frac{1}{2}}(\bar z)
\end{equation}
is also true. Invariance of $O$ with respect to other components of deformed translations follows from $\bar J$ covariance of $O$.

Equations (\ref{19sep14}) and (\ref{14sep9}) entail
\begin{equation}
\label{14sep10}
B\phi^{-\frac{1}{2}}(\bar z)=- A d \frac{i}{2} (-2)\int d\bar z_1 (\bar z-\bar z_1)^{-\frac{1}{2}}d \frac{i}{2} \int d\bar z_2 (\bar z_1-\bar z_2)^{-\frac{3}{2}}
\phi^{-\frac{1}{2}}(\bar z_2)d\bar z_2,
\end{equation}
which, considering that
\begin{equation}
\label{14sep11}
\int d\bar z_1 (\bar z-\bar z_1)^{-\frac{1}{2}}(\bar z_1-\bar z_2)^{-\frac{3}{2}}=4\pi \delta(\bar z-\bar z_2),
\end{equation}
leads to
\begin{equation}
\label{14sep12}
B=-2\pi d^2 A.
\end{equation}
Employing (\ref{19sep12}), we find 
\begin{equation}
\label{14sep13}
d^2 = -\frac{i}{2\pi A^2}.
\end{equation}
This means that there are two solutions for $d$ that make (\ref{19sep14}) invariant with respect to the chiral Poincar\'{e} algebra. Accordingly, representation 
(\ref{19sep6}), (\ref{19sep11}), (\ref{19sep12}) can be decomposed into two invariant subspaces.

\bibliography{singleton}
\bibliographystyle{JHEP}

\end{document}